\documentclass[
reprint,
superscriptaddress,
amsmath,amssymb,
aps,
prb,
floatfix,
]{revtex4-2}
\usepackage{siunitx}
\usepackage{bm}
\usepackage{graphicx}
\usepackage{dcolumn}
\usepackage{bm}
\usepackage{xcolor} 
\definecolor{crossCyan}{HTML}{00FFFF}  
\usepackage[colorlinks=true,linkcolor=blue,citecolor=blue,urlcolor=blue]{hyperref}

\begin{document}


\title{Electrical driving of hole spin states in planar silicon MOS device by \textit{g}-matrix modulation}

\author{Aaquib Shamim}
\email{a.shamim@unsw.edu.au}
\affiliation{School of Physics, University of New South Wales, Sydney NSW 2052, Australia}

\author{Scott D. Liles}
\email{s.liles@unsw.edu.au}
\affiliation{School of Physics, University of New South Wales, Sydney NSW 2052, Australia}

\author{Joe Hillier}
\affiliation{School of Physics, University of New South Wales, Sydney NSW 2052, Australia}
\author{Jonathan Y. Huang}
\affiliation{School of Physics, University of New South Wales, Sydney NSW 2052, Australia}

\author{Isaac Vorreiter}
\affiliation{School of Physics, University of New South Wales, Sydney NSW 2052, Australia}

\author{Pratik Chowdhury}
\affiliation{School of Physics, University of New South Wales, Sydney NSW 2052, Australia}

\author{Chris C. Escott}
\affiliation{School of Electrical Engineering and Telecommunications, University of New South Wales, Sydney NSW 2052, Australia}
\affiliation{Diraq, Sydney, NSW, Australia}

\author{Fay E. Hudson}
\affiliation{School of Electrical Engineering and Telecommunications, University of New South Wales, Sydney NSW 2052, Australia}
\affiliation{Diraq, Sydney, NSW, Australia}

\author{Wee Han Lim}
\affiliation{School of Electrical Engineering and Telecommunications, University of New South Wales, Sydney NSW 2052, Australia}
\affiliation{Diraq, Sydney, NSW, Australia}

\author{Kok Wai Chan}
\affiliation{School of Electrical Engineering and Telecommunications, University of New South Wales, Sydney NSW 2052, Australia}
\affiliation{Diraq, Sydney, NSW, Australia}

\author{Rajib Rahman}
\affiliation{School of Physics, University of New South Wales, Sydney NSW 2052, Australia}

\author{Andrew S. Dzurak}
\affiliation{School of Electrical Engineering and Telecommunications, University of New South Wales, Sydney NSW 2052, Australia}
\affiliation{Diraq, Sydney, NSW, Australia}

\author{Alexander R. Hamilton}
\email{alex.hamilton@unsw.edu.au}
\affiliation{School of Physics, University of New South Wales, Sydney NSW 2052, Australia}


\begin{abstract}
Hole spins in group IV quantum dots are a highly promising way to develop CMOS compatible spin qubits owing to their inherent spin-orbit coupling, which enables fast, coherent, and electrical spin control. However, spin–orbit coupling not only enables multiple spin-control mechanisms, but also exposes the qubits to charge noise. In this work, we perform a systematic study of the spin control mechanism in a planar silicon hole quantum dot. 
We use $g$-matrix formalism to discern contributions from the various spin driving mechanisms and identify regions where spins are less sensitive to charge noise. By mapping out the dependence of the Rabi frequency on the magnetic field orientation, we observe the largest Rabi frequency in the in-plane direction and the smallest Rabi frequency close to the out-of-plane direction. These results enhance the understanding of how different mechanisms contribute to spin driving within an industrially relevant architecture and aid in establishing the operating conditions for the rapid and coherent manipulation of hole qubits.

\end{abstract}

\maketitle


\section{Introduction}
The spin states of electrons and holes confined in semiconductor quantum dots serve as a building block for encoding quantum information~\cite{Loss1998}. Silicon metal-oxide-semiconductor (MOS) quantum dots~\cite{Angus2007} have emerged as a promising platform for storing and manipulating quantum information due to their compatibility with standard CMOS manufacturing processes. Silicon spin qubits have gained prominence due to long coherence times~\cite{Veldhorst2014}, the demonstration of two-qubit logic gate operations~\cite{Veldhorst2015}, and compact device architectures~\cite{compactdevice_Veldhorst2017}. Among the different device architectures, the planar silicon CMOS structure allows for dense arrays of interconnected qubits~\cite{Siinterconnected6qubits,Li2018,Li2025TrilinearQD}.

Spin qubits can be realized using electron spins~\cite{Veldhorst2014,Fogarty2018,Takeda2016} or hole spins~\cite{Maurand2016,Watzinger2018,Hendrickx2020}. For electron spin qubits, spin control is achieved using  micromagnets~\cite{Takeda2016,micromagnet_two_axis_control_Wu2014,Leon2020} or ESR strip lines~\cite{Fogarty2018,Veldhorst2015}.
In contrast, hole spin qubits benefit from an inherent spin-orbit coupling (SOC) arising from the p-type nature of holes, allowing all-electrical spin control via electric dipole spin resonance (EDSR)~\cite{Bulaev2007}. The inherent SOC removes the need for additional fabrication of micro-magnets or ESR strip lines. Furthermore, both the SOC strength~\cite{so_tuning_Froning2021} and the $g$-factor are electrically tunable for hole spins ~\cite{NAres2013_gafctortunability,Liles2021,Crippa2018}. An additional advantage of hole qubits is their weak hyperfine interaction~\cite{Prechtel2016}, which contributes to enhanced coherence times without needing to isotopically enrich Si~\cite{JoeIk2024,Liles2024STQubit}. Recent work has shown that it is possible to achieve greater than $\SI{99}{\percent}$ fidelity for holes in natural Si~\cite{Vorreiter2025}. Moreover, the tunability of the SOC can be leveraged to access decoherence ``sweet-spots'', further improving the performance of the qubits~\cite{Piot2022,Carballido2025,Wang2021,Bassi2024}. 
%


For fast spin driving, the understanding of different driving mechanisms and their interplay is essential. The mechanisms responsible for driving spins via EDSR can be decomposed into two components: $g$-tensor magnetic resonance ($g$-TMR)~\cite{Kato2003} and iso-Zeeman (IZ) ~\cite{VenitucciNiquet2018,Crippa2018}. Their interplay can be modeled using the $g$-matrix formalism~\cite{VenitucciNiquet2018}. This provides a framework for connecting the spin driving to the standard measurements of the $g$-matrix and its voltage derivative. This formalism has previously been used to discern the different spin driving mechanisms~\cite{Crippa2018} for the optimal operation of hole spin qubits in silicon nanowire systems~\cite{Bassi2024}. However, an equivalent analysis has not yet been performed in lateral quantum dots (2D) defined in planar silicon MOS devices.



The SOC physics in planar devices differs from that in nanowires due to their different confinement geometries, electric fields, interface environments, and heavy hole and light hole mixing. This makes the hole properties unique for each system ~\cite{holes_nanowire_Zwanenburg2009,holesinMOS_Spruijtenburg2013,holesinMOSLi2015}. Consequently, SOC driven phenomena can vary significantly between the two device systems. These factors necessitate the study of spin control mechanisms of lateral quantum dots defined in planar silicon MOS. 
\begin{figure*}[!t]
    \centering
    \includegraphics[width=0.75\linewidth]{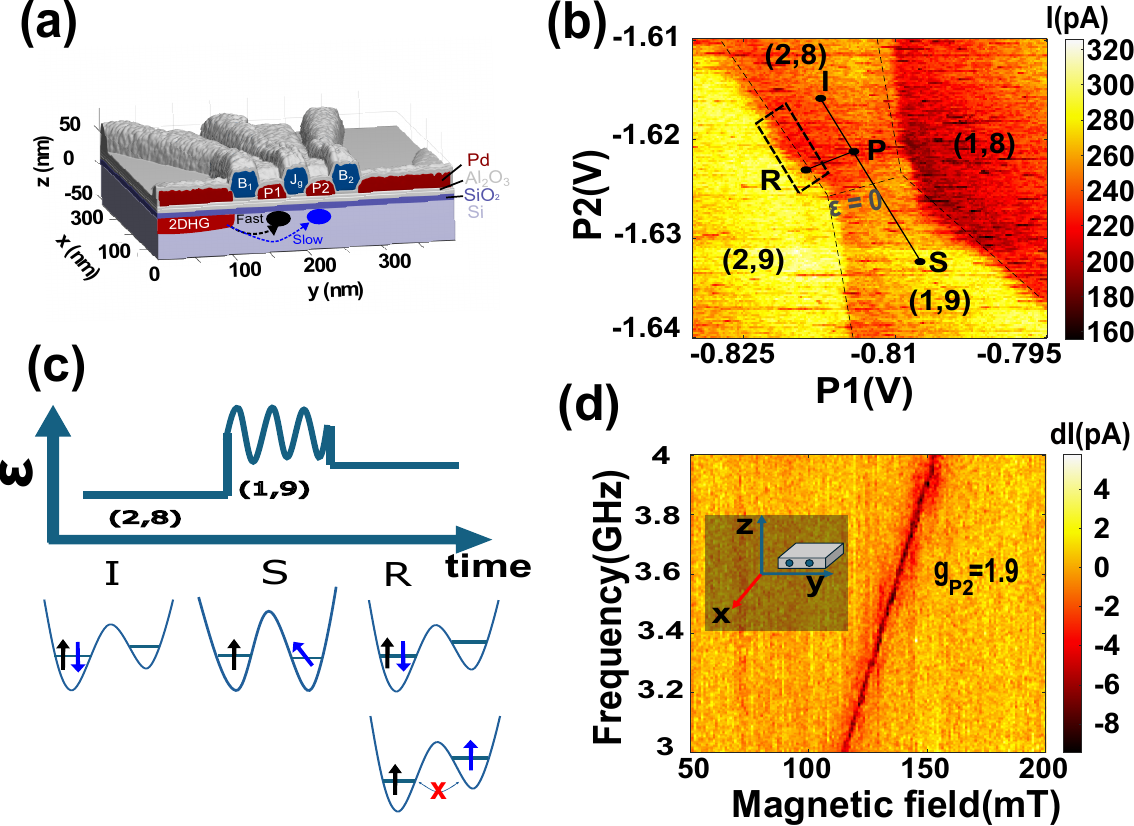}
    \renewcommand{\figurename}{FIG.}
    \caption{Device operating point and latched readout. (a) A schematic of the device showing the layout of gates and the position of the quantum dot. Hole quantum dots are formed under gates P1 and P2. The difference in tunneling rates from reservoir to the dots (fast and slow) enables latched readout. (b) Charge stability diagram of the (2,8)$\leftrightarrow$(1,9) transition. The labels indicate the numbers of holes in the P1 and P2 quantum dots. The color axis represents the electrical current $I$ through the sensor. The pulse sequence used for spin control and readout is shown in the figure. The pulse starts from the initialization point (I) in (2,8), followed by the separation point (S) which pushes the spin into (1,9). For readout, the pulse first goes to the projection point, and then go to the latched readout point (R). %
    (c) A schematic showing the pulse sequence used for spin manipulation. The figure also shows how the spins in quantum dots evolve during the pulse sequence. Initialization (I) and readout (R) are performed in the (2,8) state and manipulation (S) happens in the (1,9) state. (d) Change in sensor current ($\mathrm{d}I$) due to the pulse sequence (described above) measured as a function of microwave frequency and the magnetic field magnitude along the x-direction (shown in inset). The region where the spin resonance condition is satisfied results in enhancement of current (red line). The slope of the line gives the $g$-factor of one of the quantum dots.}
    \label{figure_1_g-TMR}
\end{figure*}

In this work, we demonstrate coherent spin control in a planar silicon MOS hole double quantum dot and provide a quantitative analysis of spin driving mechanisms. We employ the $g$‑matrix formalism to disentangle the different driving mechanisms. We find that the dominant spin‑driving mechanism in this system is the IZ contribution. The maximal Rabi frequency is in the direction perpendicular to the spin-orbit field $\bm{B_\mathrm{SO}}$, while the minimum is close to the out-of-plane direction. From the gate‑voltage dependence of the $g$‑matrix, we identify the ``sweet-spots'' for coherent operation. These measurements provide a comprehensive characterization of the electrical driving of the hole spins in a planar lateral MOS structure.

\section{Results}
\subsection{Device and operation regime}

The device studied was a hole double quantum dot defined in planar silicon, fabricated using MOS techniques~\cite{IKJin2023}. Fig.~\ref{figure_1_g-TMR}(a) shows a schematic of the device. The gates labeled P1 and P2 serve as plunger gates, while B1 and B2 act as barrier gates. The J$_\mathrm{g}$ gate controls the inter-dot tunnel coupling \textit{t}$_\mathrm{c}$ \cite{IKJin2023}. The device includes an adjacent nMOS single electron transistor (SET) which is used as a charge sensor (not shown in the figure). All measurements shown in this work are performed using the charge sensor. 
\begin{figure*}[!t]
\centering
\includegraphics[width=0.85\linewidth]{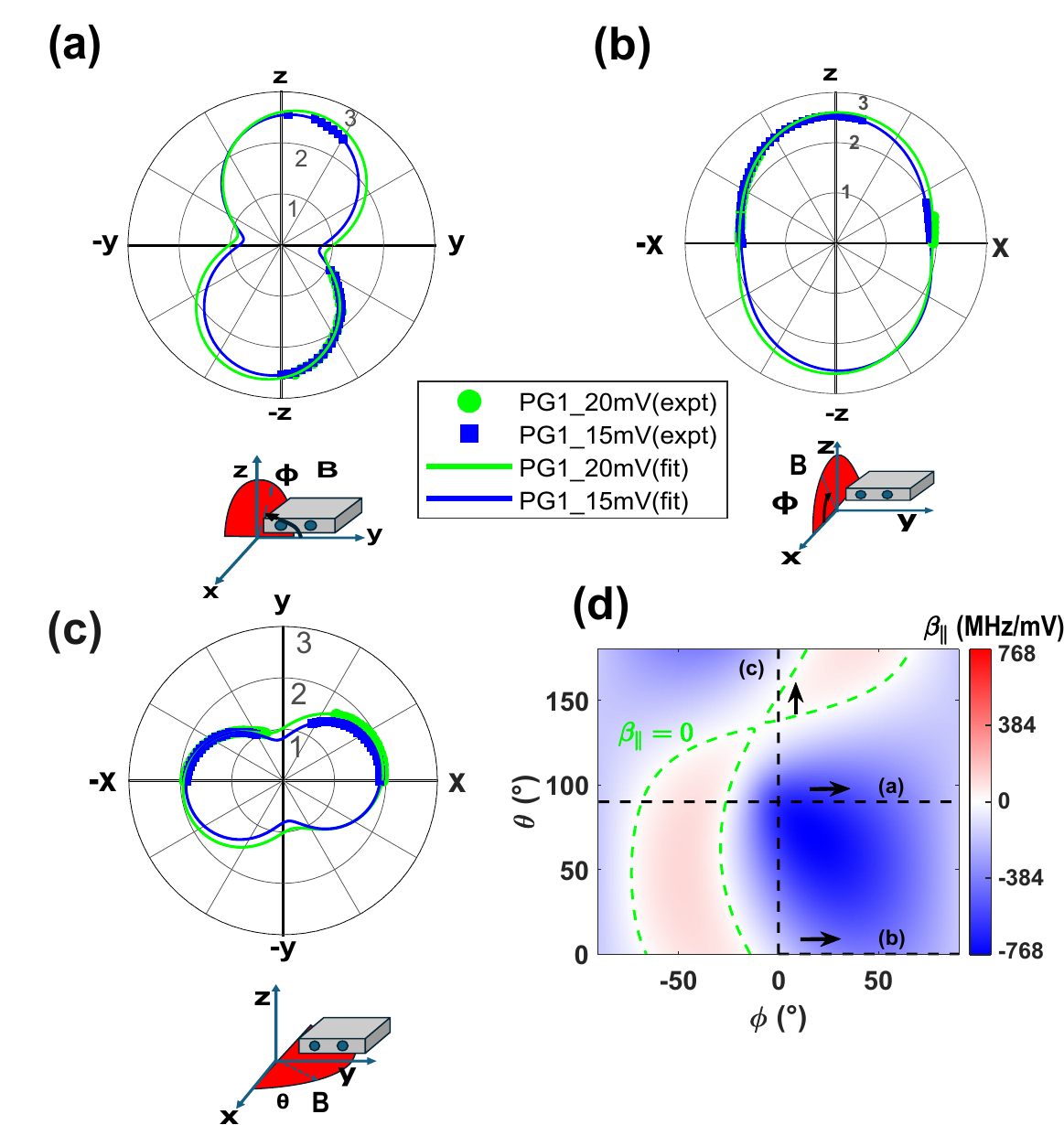}
\renewcommand{\figurename}{FIG.}
\caption{Mapping of $g$-factor anisotropy at two different P1 voltages. $g$-factor ($|g^*|$) measured for different orientation (defined by $\theta$ and $\phi$) of the magnetic field around the sample x (a), y (b), and z (c) axes. The magnetic field was kept at a constant magnitude of $\SI{0.130}{\tesla}$. Different colors indicate different P1 voltage pulses at which the experiment was performed. The markers are experimental data, and the solid line is fit to the Eq.~\ref{equation_1_g-TMR}. The schematic below each polar plot defines the plane of rotation for which the experiment is being done. (d) Angular dependence of $\beta_{||}$ as a function of angle. The green line marks the ``sweet lines'', corresponding to the region where $\beta_{||}=0$.}  
\label{figure_2_g-TMR}
\end{figure*}

Fig.~\ref{figure_1_g-TMR}(b) shows a charge stability diagram of the (2,8)$\leftrightarrow$(1,9) configuration. The color scale corresponds to the current through the sensor. We perform all measurements in the (2,8)$\leftrightarrow$(1,9) charge transition regime, which is equivalent to a (2,0)$\leftrightarrow$(1,1) spin system \cite{Liles2018}. Spins are initialized in the (2,0) equivalent state by dwelling in (2,8) state at point I. From I, the state is pulsed to position S where spin manipulation happens. For readout at point R, we  use latched Pauli-spin-blockade (PSB)~\cite{Latching2018_Patrick,Studenikin2012}, which allows us to perform spin-to-charge conversion (Appendix A). We operate at a large detuning ($\epsilon$) where the quantum dots are effectively decoupled due to reduced wavefunction overlap, resulting in a negligible exchange interaction ($J \approx 0$). In this regime, the holes are well localized in their respective dots, and the system can be treated as two individually controllable spins, as shown in Fig.~\ref{figure_1_g-TMR}(c). Fig.~\ref{figure_1_g-TMR}(c) also illustrates the detuning pulse used to perform EDSR and the spin configurations at different stages of the pulse sequence. To drive spin transitions, we apply a microwave signal to the P1 gate during stage S, allowing resonant spin control through EDSR. At point R, the spin states can either be in the singlet state or be in the blocked state. 
In Fig.~\ref{figure_1_g-TMR}(d), we measure the sensor current response as a function of the magnetic field magnitude $B$ along x and the microwave frequency $f$. The spin resonance condition is satisfied when $hf = |g^*| \mu_\mathrm{B} B$; here $h$ is Planck's constant, $\mu_\mathrm{B}$ is the Bohr magneton, and $g^*$ is the effective hole $g$-factor along the direction of the magnetic field $\bm{B}$. When the spin resonance condition is satisfied, a spin-flip transition is induced. This resonance appears as a continuous line in the map of sensor current as a function of the magnetic field and the microwave frequency. From the slope of the line, we extract the $g$-factor to be $g_\mathrm{P2} = 1.9$. Ideally, for two isolated spins, two resonance lines would be expected. Based on the angular measurements of EDSR in Figs.~\ref{figure_2_g-TMR}(a-c) we rule out the $g$-factor being the same in the two dots. From additional measurements (Appendix B), we attribute the EDSR signal to the dot under P2.

\subsection{\textit{g}-tensor for two different voltage pulses at gate P1}

In order to fully characterize the hole spin system, we extract the hole $g$-tensor ($\hat{G}$). $\hat{G}$ defines the coupling of a hole spin to the magnetic field and depends on various device factors such as crystal anisotropies, confinement profile~\cite{NAres2013_gafctortunability} and strain~\cite{Liles2021}. We extract the hole $g$-tensor by measuring the observed EDSR frequency as a function of the magnetic field orientation. The resonance frequencies are fitted with a model that relates the Zeeman splitting to the $g$-tensor:
\begin{equation}
    \Delta E^2 = |g*|^2\mu_\mathrm{B}^2 B^2 =\mu_\mathrm{B}^2|\hat{g}\cdot\bm{B}|^2 =\mu_\mathrm{B}^2(\bm{B}^\mathrm{T}\cdot \hat{G}\cdot \bm{B}),
\label{equation_1_g-TMR}
\end{equation}
where $B$ is the magnetic field, $\hat{G} = \hat{g}^\mathrm{T}\cdot\hat{g}$ is the symmetric Zeeman tensor and  $\hat{g}$ is the $g$-matrix. The eigenvalues of $\hat{G}$ are the squares of the principal $g$-factors. The eigenvectors of $\hat{G}$ are the principal magnetic axes.


Figs.~\ref{figure_2_g-TMR}(a-c) show the measured $g$-factor over a range of magnetic field orientations for the yz, xz and xy planes, respectively. We carry out measurements over an angular range of $\SI{180}{\degree}$ because the Zeeman splitting 
is invariant under field reversal. The measured data (blue and green markers) are fitted to Eq.~\ref{equation_1_g-TMR} (solid lines). We observe that there are regions with clear EDSR signals interrupted by intervals without a detectable resonance. The visibility of the EDSR signal enhances by increasing the microwave driving power, but there are still regions where there is no EDSR signal (Appendix D).

From the fit we extract $\hat{G}$ in the laboratory frame $(\mathrm{x},\mathrm{y},\mathrm{z})$ as
\begin{equation}
\hat{G} =
\begin{pmatrix}
3.56 & 0.36 & -0.26 \\
0.36 & 0.67 & 0.82 \\
-0.26 & 0.82 & 6.46
\end{pmatrix}.
\label{equation_2_g-TMR}
\end{equation}
From the square roots of the eigenvalues of $\hat{G}$, we determine the principal $g$-factors to be $|g_\mathrm{x}|^* = 1.9$, $|g_\mathrm{y}|^* = 0.71$, and $|g_\mathrm{z}|^* = 2.57$. The largest $g$-factor is slightly tilted away by $\approx \SI{15}{\degree}$ from the z-axis, which is the direction of the strongest confinement~\cite{Liles2021}.

While $\hat{G}$ captures the anisotropy of the spin system, further insight is gained from its voltage dependence, $\hat{G}'=\frac{\mathrm{d}\hat{G}}{\mathrm{d}V}$. $\hat{G}'$ governs the spin driving contribution due to $g_\mathrm{TMR}$ and quantifies the longitudinal electric susceptibility $\beta_{||}$, which couples the qubit to charge noise \cite{Bassi2024}. Ideally the qubit operation point should be where the $\beta_{||}$ is minimized, at the so called ``sweet-spots''. $\beta_{||}$ is related to $\hat{G}'$ using
\begin{equation}
    \beta_{||} =\frac{1}{2h\Delta E} {\mu _\mathrm{B}^2\bm{B}^\mathrm{T}\cdot \hat{G}'\cdot\bm{B}}.
\label{equation_3_g-TMR}
\end{equation}
To determine $\hat{G}'$, we calculate $\hat{G}$ at two different P1 voltages to extract the voltage dependence of the $g$-tensor. We focus on P1 because it is the driving gate and therefore directly contributes to the Rabi frequency. We measure $\hat{G}'$ in the laboratory frame $(\mathrm{x},\mathrm{y},\mathrm{z})$ to be
\begin{equation}
\hat{G}' = 
\begin{pmatrix}
-33.24 & -51.05 & -73.51 \\
-51.05 & -72.74 & -88.13 \\
-73.51 &  -88.13 & -58.59
\end{pmatrix}.
\label{equation_4_g-TMR}
\end{equation}
Previous work \cite{Bassi2024,NicoHendrickx2024,Carballido2025} has shown that field orientations where $\hat{G}'$ vanish correspond to maximal coherence, as the coupling to charge noise is minimized. For the device under study, we calculate $\beta_{||}$ and use Eq.~\ref{equation_3_g-TMR} to identify the ``sweet-lines'' \cite{Bassi2024}, where $\frac{\mathrm{d}G}{\mathrm{d}V_\mathrm{g}}$ is minimized. 
\begin{figure*}[!t]
\centering
\includegraphics[width=0.75\linewidth]{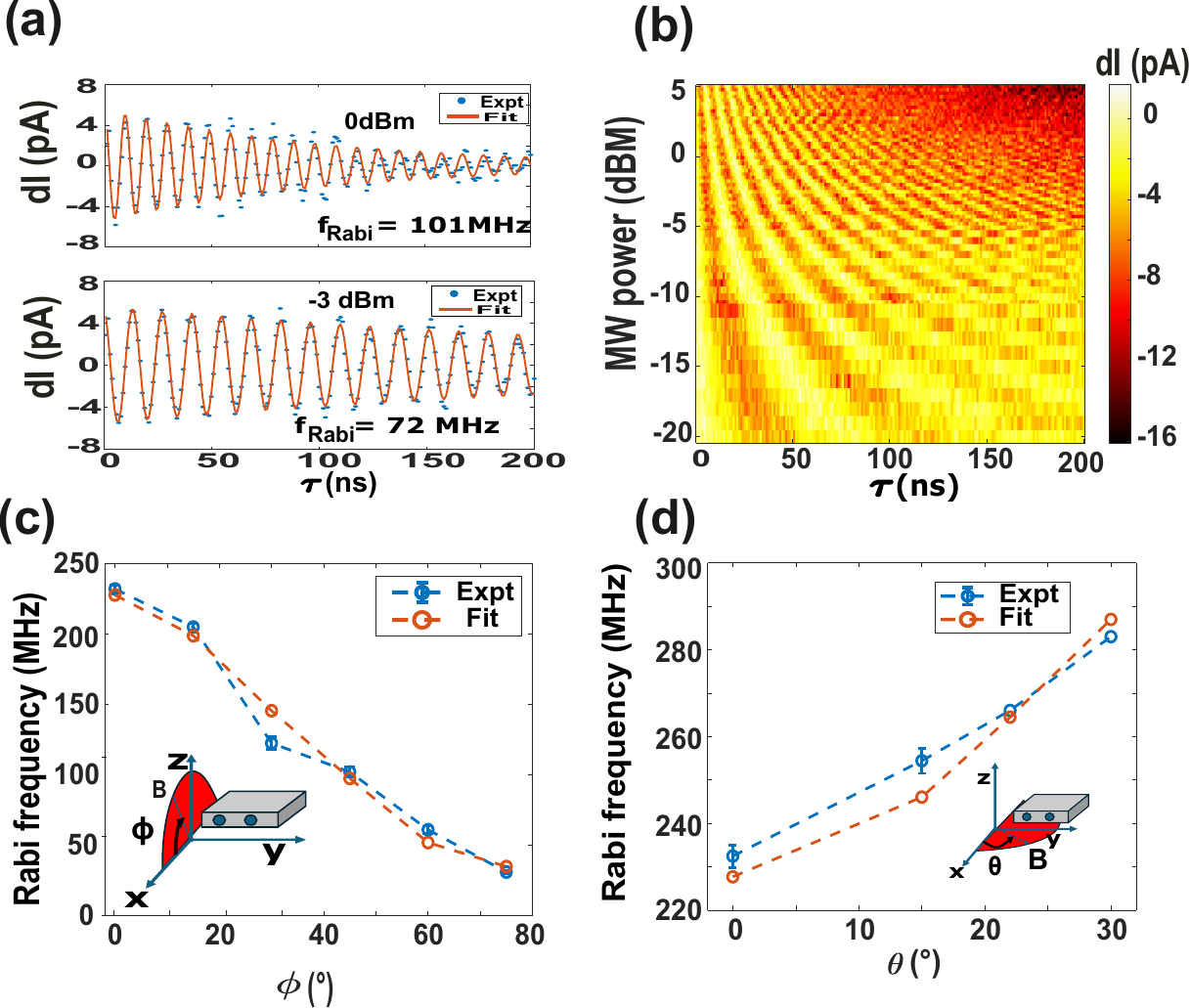}
\renewcommand{\figurename}{FIG.}
\caption{Coherent spin manipulation and anisotropy in Rabi frequency. (a) Rabi oscillations for different levels of driving power. The markers are experimental data and the solid lines are the fits to the equation $A\sin (2\pi f_\mathrm{Rabi}\tau) e^{-\frac{\tau}{d_1}}$. Rabi frequencies are $\SI{101}{\mega\hertz}$ and $\SI{72}{\mega\hertz}$ for microwave power of $\SI{0}{dBm}$ and $\SI{-3}{dBm}$ respectively. Magnetic field of magnitude $\SI{0.135}{\tesla}$ is applied in the x-direction. The microwave frequency is $\SI{3.5}{\giga\hertz}$. (b) Color plot of Rabi oscillations for a range of microwave power and burst time at the same microwave frequency and magnetic field magnitude as in (a). (c)-(d) Rabi frequency as a function of the magnetic field orientation at microwave power of $\SI{5}{dBm}$. The blue markers are the data point and orange markers are fitted to Eq.~\ref{equation_5_g-TMR}. The inset schematic shows the direction of the plane of rotation. The error bars for $\theta=\SI{22}{\degree}$ and $\SI{30}{\degree}$ in (d) are not obtained as the Rabi frequency values are calculated using FFT (Fast Fourier Transform) of the time domain data instead of curve fitting.}
\label{figure_3_g-TMR}
\end{figure*}

These $\beta_{||}$ ``sweet-lines'' provide insight into regions of reduced charge noise induced dephasing without requiring a direct $T_2^*$ measurement. Fig.~\ref{figure_2_g-TMR}(d) shows $\beta_{||}$ as a function of the magnetic field orientation. The green lines in the figure mark the ``sweet-lines''. The regions with enhanced $\beta_{||}$ are the regions of maximal dephasing due to charge noise. The black dotted lines with labels indicate the planes where the experiments in Figs.~\ref{figure_2_g-TMR}(a-c) were performed. Notably, the orientations where the $g$-factor data could not be acquired in Figs.~\ref{figure_2_g-TMR}(a–c) coincide with regions of enhanced $\beta_{||}$. For example, in Fig.~\ref{figure_2_g-TMR}(d) along the black dotted line labeled (c), which corresponds to the plane in Fig.~\ref{figure_2_g-TMR}(c), $\beta_{||}$ evolves from nearly zero to its maximum and then decreases again. Consistently, in Fig.~\ref{figure_2_g-TMR}(c), $g$-factor data is not acquired around the region of enhanced $\beta_{||}$. A similar relation can be observed by comparing Figs.~\ref{figure_2_g-TMR}(a-b) with the lines labeled (a) and (b) in Fig.~\ref{figure_2_g-TMR}(d). Thus, the missing $g$-factor data points can be attributed to increased sensitivity to charge noise (large $\beta_{||}$).   

\subsection{Rabi frequency anisotropy and the contribution of different mechanisms}
Having characterized the anisotropic spin response through the full $g$-tensor and identified the ``sweet-spots'', we now turn to dynamic measurements of coherent spin control via Rabi oscillations. We characterize the orientation dependence of the Rabi frequency. These measurements not only quantify the spin rotation rates but also provide a means to disentangle the relative contributions of IZ and $g$‐TMR contributions to electrical spin driving.

Fig.~\ref{figure_3_g-TMR}(a) shows Rabi oscillations as a function of microwave burst time $\tau$ at two different levels of microwave power ($\SI{0}{dBm}$ and $\SI{-3}{dBm}$). The higher the microwave power, the faster the spin is driven. Fig.~\ref{figure_3_g-TMR}(b) presents a colormap of Rabi oscillations for a range of microwave powers and burst time. To understand the underlying mechanisms driving these oscillations, we use the following expression for the Rabi frequency\cite{Crippa2018,Ares2013}: 
\begin{equation}
f_\mathrm{Rabi}=\frac{\mu_\mathrm{B} B V_\mathrm{ac}}{2h|g^*|}[\hat{g}(V_0)\cdot\hat{b}] \times [\hat{g}'(V_0)\cdot\hat{b}],
\label{equation_5_g-TMR}
\end{equation}
where $\hat{g}'$ is the $g$-matrix derivative with respect to the gate voltage, $\hat{b}$ is the unit vector along the direction of the magnetic field, $B$ is the magnetic field magnitude, and $V_\mathrm{ac}$ is the amplitude of the driving signal. While the Zeeman splitting is governed by the symmetric tensor $\hat{G} = \hat{g}^\mathrm{T}\cdot\hat{g}$, the Rabi frequency depends on the cross product of $\hat{g}\cdot\hat{b}$ and $\hat{g}'\cdot\hat{b}$ . This form captures how spin rotations arise from the misalignment between the static spin quantization axis and the electrically driven axis. To identify the dominant driving mechanism, we measured the angular dependence of the Rabi frequency, which enables us to disentangle and quantify the relative contributions from $g$-TMR and IZ. The contributions of $g$-TMR and IZ are expressed as $\hat{g}'_\mathrm{TMR}$ and $\hat{g}'_\mathrm{IZ}$ respectively.

Fig.~\ref{figure_3_g-TMR}(c) and ~\ref{figure_3_g-TMR}(d) show the measured and fitted Rabi frequency using Eq.~\ref{equation_5_g-TMR} for different angles with $\hat{g}'_\mathrm{IZ}$ and amplitude of the microwave tone ($V_\mathrm{ac}$) as fitting parameters (discussed in Appendix C). The schematic inset indicates the measurement plane. For each measurement, the Zeeman splitting is fixed at $\SI{3.6}{\giga\hertz}$ by adjusting the magnetic field to compensate for the frequency-dependent attenuation in the RF lines.
\begin{figure*}
    \centering
    \includegraphics[width=0.8\linewidth]{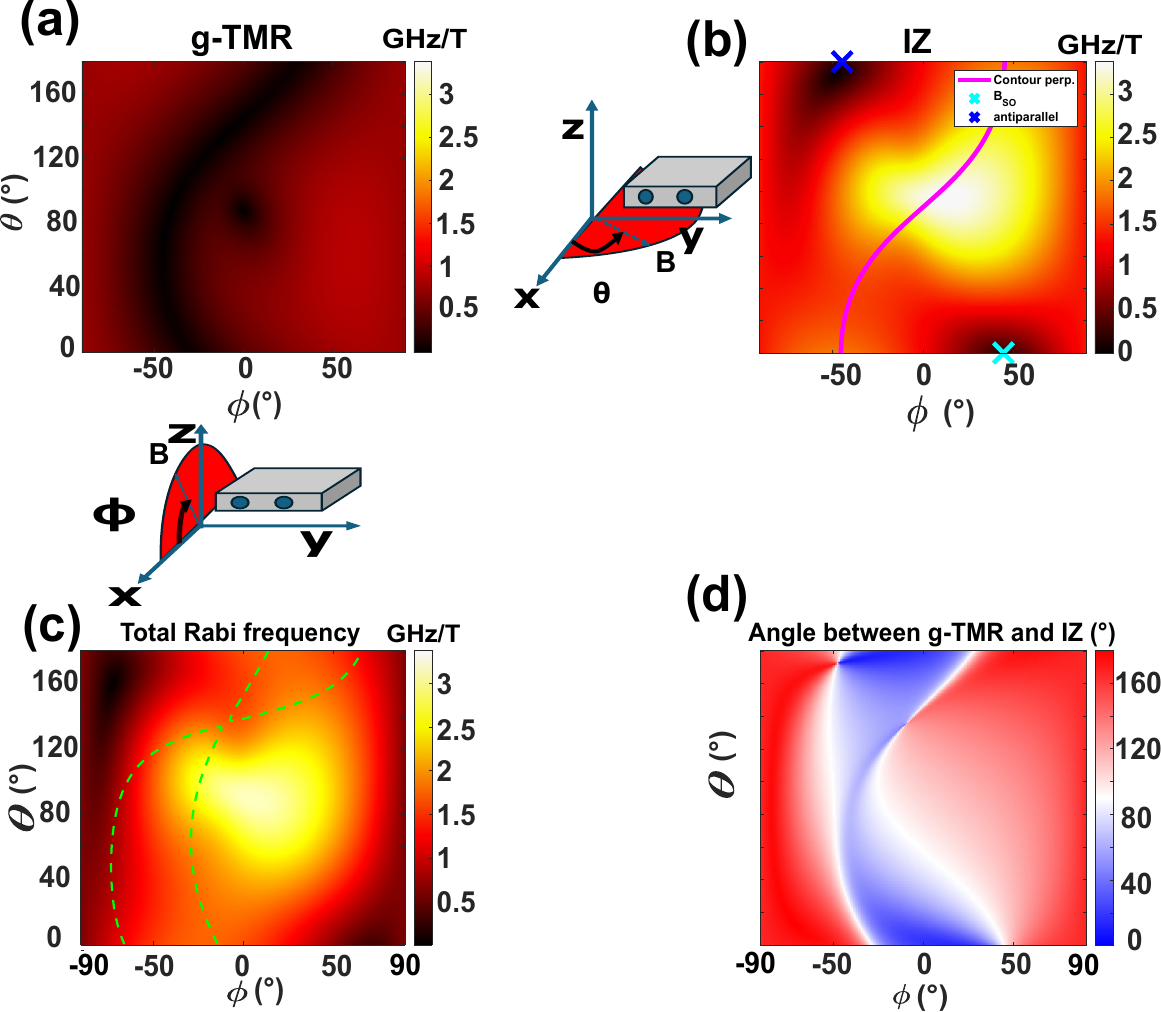}
    \renewcommand{\figurename}{FIG.}
    \caption{Spin driving mechanism for Rabi drive. (a) $g$-TMR and (b) IZ contributions to Rabi frequency extracted from fitting. The ``$\textcolor{crossCyan}{\times}$'' is the chosen $\bm{B_\mathrm{SO}}$ direction where a minimum in Rabi frequency occurs. The ``$\textcolor{blue}{\times}$'' is same as  ``$\textcolor{crossCyan}{\times}$'' rotated by $\SI{180}{\degree}$. The continuous ``$\textcolor{pink}{\rule[0.35ex]{2.4ex}{0.5ex}}$'' line includes all possible directions which are perpendicular to ``$\textcolor{crossCyan}{\times}$''(``$\textcolor{blue}{\times}$'').   
    (c) Total Rabi frequency with enhanced coherence sweetlines ($\beta_{||}=0$) as dashed green lines reproduced from Fig.~\ref{figure_2_g-TMR}. (d) Angle between the $g$-TMR and IZ components. 
    }
    \label{figure_4_g-TMR}
    \end{figure*}
\noindent The Rabi frequency varies from $\SI{290}{\mega\hertz}$ ( in-plane direction) to $\SI{20}{\mega\hertz}$ (out-of-plane direction: $\phi = \SI{75}{\degree}$), highlighting the strong anisotropy in the spin driving. Fitting Eq.~\ref{equation_5_g-TMR} to the experimental data in Figs.~\ref{figure_3_g-TMR}(c-d) and following the procedure described in Ref.~\cite{Crippa2018}, we determine $\hat{g}'_\mathrm{TMR}$ and $\hat{g}'_\mathrm{IZ}$ to be:
\begin{equation}
\begin{aligned}
\hat{g}'_\mathrm{TMR} & =
\begin{pmatrix}
-7.13 & -10.95 & -15.82\\
-28.04 &- 40.93 & -52.34\\
-11.91 & -13.61 & -6.75\\
\end{pmatrix},\\
\hat{g}'_\mathrm{IZR} & = 
\begin{pmatrix}
-23.53 & 217.66 &  47.67\\
-86.64 & 39.19 &  83.62\\ 
26.99 & 73.96 & -15.67
\end{pmatrix}.
\end{aligned}
\label{equation_6_g-TMR}
\end{equation}

We use the fitted quantities to reconstruct the full behavior of the Rabi frequency as a function of the applied magnetic field orientation. We use this analysis to gain a deeper understanding of the interplay between the different spin driving mechanisms. 

To begin with, we focus only on the contribution of the $g$-TMR driving mechanism. $g$-TMR is associated with the change in shape of the wave function and the confinement potential of the dot. This modulates the $g$-factor and hence the Larmor vector, enabling spin rotations when the system is driven at resonance.  Fig.~\ref{figure_4_g-TMR}(a) shows the $g$-TMR contribution to the spin driving as a function of the applied magnetic field orientation. This contribution is calculated from Eq.~\ref{equation_5_g-TMR} using $\hat{g}'_\mathrm{TMR}$ matrix in Eq.~\ref{equation_6_g-TMR} as an input for $\hat{g}'$ term. We observe a maximal $g$-TMR rate of $\SI{0.91}{\giga\hertz/\tesla}$, while the minima follow a curved contour. Based on this, we can identify the applied magnetic field directions that maximize or minimize the $g$-TMR driving.

We can understand the $g$-TMR response to spin driving by analyzing how the key parameters in Eq.~\ref{equation_5_g-TMR} interact with one another ($\hat{g}$, $\hat{g}'_\mathrm{TMR}$ and $\hat{b}$). There are three conditions that will lead to minima in the $g$-TMR Rabi frequency; (1) when the magnitude of $\hat{g}\cdot\hat{b}$ is minimized; (2) when the magnitude of $\hat{g}'_\mathrm{TMR}\cdot\hat{b}$ is minimized; (3) when the angle between the vectors $\hat{g}'_\mathrm{TMR}\cdot\hat{b}$ and $\hat{g}\cdot\hat{b}$ is zero (the cross product term vanishes). By analyzing the functional forms of our parameters (Appendix E), we attribute the small $g$-TMR minimum region near $(\phi,\theta)\approx(0,80)$ to the vanishing of the cross product term in Eq.~\ref{equation_5_g-TMR} approaching zero. We ascribe the minimal region following the continuous curved ridge to minima in the magnitude of the $\hat{g}'_\mathrm{TMR}\cdot\hat{b}$. We can also understand the maxima in the $g$-TMR contribution as arising from the maxima in the magnitude of $\hat{g}'_\mathrm{TMR}\cdot\hat{b}$.

Next, we focus on the contribution of the IZ driving mechanism which arises as the applied RF field oscillates the wave function. Due to $\bm{B_\mathrm{SO}}$, this motion gives rise to an effective time-dependent magnetic field. This oscillating magnetic field, when matched with the Zeeman splitting, drives spin rotations. Since there is no modulation of the Zeeman energy, the contribution from this mechanism can only be determined from time-domain measurements of the Rabi frequency as a function of magnetic field orientation. Fig.~\ref{figure_4_g-TMR}(b) shows the IZ contribution to the spin driving as a function of the orientation of the applied magnetic field. The IZ contribution can be calculated using the $\hat{g}'_\mathrm{IZ}$ matrix as the input to Eq.~\ref{equation_5_g-TMR}. We observe a maximal IZ rate of $\SI{3.3}{\giga\hertz/\tesla}$, which occurs when the magnetic field is close to the sample y-axis - ($\phi,\theta$) $\approx$ $(0,\SI{90}{\degree})$. The maximal IZ contribution exceeds the maximal $g$-TMR contribution by more than a factor of three, indicating that the IZ mechanism dominates, consistent with holes in silicon nanowires \cite{Crippa2018, Carballido2025}. We observe two minima in the IZ map, which are marked with colored crosses ``$\textcolor{crossCyan}{\times}$'' and ``$\textcolor{blue}{\times}$''. These two minima correspond to anti-parallel vectors, i.e. they are $\SI{180}{\degree}$ apart.

We can understand the IZ response in terms of the spin-orbit vector and the form of the $\hat{g}'_\mathrm{IZ}$ term. The IZ contribution comes from the motion of the dot under the oscillating microwave field, which gives rise to $\bm{B_\mathrm{SO}}$. When the externally applied magnetic field ($\bm{B}$) is parallel or antiparallel to the $\bm{B_\mathrm{SO}}$ vector, the IZ contribution to the Rabi frequency will be minimized. We see two minima in the IZ map (Fig.~\ref{figure_4_g-TMR}(b)) at $\theta = 0$, $\phi = \SI{43}{\degree}$ (``$\textcolor{crossCyan}{\times}$'') and $\theta = \SI{180}{\degree}$, $\phi = -\SI{43}{\degree}$(``$\textcolor{blue}{\times}$''). We understand these two minima as corresponding to the direction of $\bm{B_\mathrm{SO}}$ vector (parallel/antiparallel to $\bm{B}$). Therefore, from these measurements, we are able to identify the orientation of the spin-orbit vector, and interestingly we observe that there is an out-of-plane component of $\bm{B_\mathrm{SO}}$ which cannot be explained by Rashba or Dresselhaus SOC~\cite{Winkler2003}.  

The IZ component is expected to be maximal for any applied $\bm{B}$ which is perpendicular to $\bm{B_\mathrm{SO}}$~\cite{ZhanningWang2024}. The continuous pink line in Fig.~\ref{figure_4_g-TMR}(b) is the set of vectors which are perpendicular to $\bm{B_\mathrm{SO}}$. While the IZ contribution is additionally complicated by anisotropy in the $\hat{g}\cdot\hat{b}$ term, we find good agreement that all points along the pink line tend towards local maxima in Rabi frequency. Further, the global maximum in Rabi frequency due to IZ occurs at $\phi = \SI{14}{\degree}$, $\theta = \SI{95}{\degree}$, which falls on the pink line. 


We can roughly estimate the direction of the wavefunction motion in this device based on the analysis of the IZ component of spin driving. For holes in silicon, we expect the SOC to be Rashba type \cite{Winkler2003}. This Rashba SOC results in a $\bm{B_\mathrm{SO}}$ that is perpendicular to the direction of motion in the x-y plane (e.g., motion along the y-axis would give a $\bm{B_\mathrm{SO}}$ along the x-axis). As a result, we can expect that the IZ driving will be maximized when the applied magnetic field is aligned with the direction of the wavefunction motion \cite{ZhanningWang2024}. In Fig.~\ref{figure_4_g-TMR}(b), the in-plane section ($\phi$=0) of the IZ driving has a maximum for a magnetic field aligned approximately along the y-axis ($\theta \approx \SI{90}{\degree}$). This suggests that the direction of the wavefunction motion is along the sample y-axis. Additionally, we comment that we are using the P1 gate to drive the dot under P2, which qualitatively supports driving along the y-axis. However, in Fig.~\ref{figure_4_g-TMR}(b) we observe that the $\bm{B_\mathrm{SO}}$ vector has an out-of-plane component. This out-of-plane component of $\bm{B_\mathrm{SO}}$ cannot be fully explained by the simple 2D Rashba SOC, suggesting that additional forms of SOC may be needed to fully explain the spin driving mechanism.

Finally, we investigate the overall Rabi frequency dependence on the magnetic field orientation. In Fig.~\ref{figure_4_g-TMR}(c), we plot the total Rabi frequency, including both the $\hat{g}'_\mathrm{TMR}$ and the $\hat{g}'_\mathrm{IZ}$ terms in Eq.~\ref{equation_5_g-TMR}. The overall trend is similar to the trend of the IZ component in Fig.~\ref{figure_4_g-TMR}(b). The magnitude of the IZ component tends to dominate the $g$-TMR components. The maximal overall Rabi frequency occurs when $\bm{B}$ is applied approximately along the sample y-axis -- $(\phi,\theta)\approx(0,90)$. Interestingly, this point corresponds to a maximum in the IZ component and a minimum in the $g$-TMR component, indicating that the maximal Rabi driving occurs with almost purely IZ components. An attractive configuration for spin orbit qubits are ``sweet-spots'', where both coherence and driving speed are maximized. For reference, we plot the enhanced coherence ``sweet-lines'' ($\beta=0$ from Fig.~\ref{figure_2_g-TMR}(d)) as dashed green lines on the overall Rabi map. The potential optimal operation point of this qubit can be identified as a point along the ``sweet-lines'' which has the largest Rabi frequency and the longest coherence. For this experiment the ``sweet-spot'' occurs around ($\phi ,\theta$)$\approx$ ($-\SI{15}{\degree}, \SI{100}{\degree}$). The overall trend in Fig.~\ref{figure_4_g-TMR}(c) demonstrates the total Rabi driving behavior for a planar hole spin qubit.

When evaluating the overall Rabi frequency ($f_\mathrm{R}$), the $g$-TMR ($f_\mathrm{TMR}$) and the IZ ($f_\mathrm{IZ}$) components are not simply additive ($f_\mathrm{R}\neq f_\mathrm{TMR}+f_\mathrm{IZ}$), but can interfere constructively or destructively. The interplay of the competing $\hat{g}'_\mathrm{IZ}$ and $\hat{g}'_\mathrm{TMR}$ terms explains why we see pronounced minima in Rabi frequency around the left and right edges of Fig.~\ref{figure_4_g-TMR}(c), despite both the IZ and $g$-TMR components not being minimal here. To illustrate this effect, Fig.~\ref{figure_4_g-TMR}(d) shows the relative angle between the $g$‑TMR and IZ components as a function of magnetic field orientation. Along the directions of the minima of total Rabi frequency (Fig.~\ref{figure_4_g-TMR}(c)), the $g$-TMR (Fig.~\ref{figure_4_g-TMR}(a)) and IZ (Fig.~\ref{figure_4_g-TMR}(b)) contributions are of similar magnitude but are antiparallel to each other. Hence, the total Rabi frequency is reduced due to the destructive interference arising from the nearly $\SI{180}{\degree}$ angle between the driving vectors. Together, Fig.~\ref{figure_4_g-TMR}(c) and Fig.~\ref{figure_4_g-TMR}(d) show that the total Rabi driving is governed by both the magnitude and the relative orientation of the $g$-TMR and IZ contributions.

 

\subsection{Conclusion}
In summary, we demonstrate coherent control of a hole spin confined in a planar silicon MOS double quantum dot, provide a quantitative analysis of its electrical driving mechanisms, and identify the regions of minimal coherence due to the charge noise. By extracting the full $g$-tensor and its voltage dependence, we reveal a strongly anisotropic spin response, both in Zeeman splitting and Rabi driving. Our results indicate that spin control in planar silicon hole spin qubits is predominantly governed by the IZ mechanism, with the $g$-TMR contribution playing a secondary role. From the IZ contribution, we are able to clearly identify the effective spin orbit field direction and estimate the physical driving direction of the dot. Unexpectedly, we observe out-of-plane components of the $\bm{B_\mathrm{SO}}$ vector, which are not typically expected from 2D spin orbit mechanism such as Rashba and Dresselhaus~\cite{Winkler2003}. The out-of-plane components of $\bm{B_\mathrm{SO}}$ may arise due to strain or disorder at the oxide interface. Furthermore, we show that the total Rabi frequency is not simply additive but is dependent on the relative orientation between the IZ and $g$-TMR driving vectors. In particular, destructive interference between the two mechanisms explains the observed suppression of Rabi frequency along specific magnetic field orientations. The identified ``sweet-spots'' reduce sensitivity to charge noise and align with some regions of maximal driving. These results provide a comprehensive understanding of spin–electric coupling in planar hole spin systems and offer guidance for optimizing spin control in scalable, CMOS‑compatible architectures.

\section{Methods}
\subsection{Sample details}
The device is fabricated on a high-resistivity silicon substrate with natural isotopic composition. 
A thermally grown SiO$_2$ gate oxide with a thickness of $\SI{5.9}{\nano\meter}$ forms the Si/oxide interface. The regions  p$^+$ and n$^+$ are doped with boron and phosphorus, respectively. 
An additional layer of $\SI{2}{\nano\meter}$ Al$_2$O$_3$ is deposited on top of SiO$_2$ to enhance gate coupling and electrically isolate the metallic gates. The gate electrodes are fabricated from Pd.
   
\subsection{Experimental setup}
The experiments are carried out in a BlueFors XLD dilution fridge equipped with a three-axis vector magnet (American Magnetics Inc). All experiments are carried out at the base temperature. The device is mounted on a brass sample holder that is thermally anchored to the cold finger of the dilution refrigerator. 
Mechanical fixation is achieved using GE varnish, while electrical connections between the device and a custom made printed circuit board are established via Al wire bonds. 
All DC gate and bias voltages are supplied by a Delft IVVI digital-to-analog converter. 
Each DC line is filtered by an individual $\SI{50}{\kilo\hertz}$ low-pass filter thermally anchored at the cold finger. 
The current through the single electron transistor (SET) charge sensor is amplified using a Basel SP983c I-V current preamplifier. 
Resulting DC currents are measured with a Keithley 2000 multimeter, and standard low-frequency lock-in measurements are performed using an SR830. Fast voltage pulses and microwave excitation are applied using a Tabor WX1284C and a Rohde \& Schwarz SGS100A, respectively.

\section{Acknowledgments}
All authors acknowledge funding from the Australian Research Council (Grants No. DP200100147 and No. FL190100167) and the US Army Research Office (Grant No. W911NF-23-1-0092). A.R.H. acknowledges an ARC Industry Laureate Fellowship (IL230100072). A.S. acknowledges funding from Sydney Quantum Academy. S.D.L acknowledges funding from ARC Early Career Industry Fellowship (IE250100140) co-funded by Diraq Pty Ltd. K.W.C. acknowledges funding from ARC Mid-Career Industry Fellowship (IM230100396). Devices were made at the New South Wales node of the Australian National Fabrication Facility.

\section{Author contributions}
A.S. and S.D.L performed all experiments and analysis. F.E.H, W.H.L. and K.W.C. fabricated the device. J.Y.H. and C.C.E. produced the 3D model Fig.~\ref{figure_1_g-TMR}(a). P.C. and R.R., contributed to the theoretical analysis and interpretation. A.S. wrote the manuscript with input from all co-authors. All authors, including including J.H., I.V., A.S.D. contributed to the discussion and planning. S.D.L and A.R.H. supervised the project.

\section{Competing interest}
A.S.D. is the CEO and director of Diraq Pty Ltd. C.C.E., F.E.H., W.H.L., K.W.C and A.S.D. declare equity interest in Diraq Pty Ltd. The remaining authors declare no competing interests.
\appendix

\section{\NoCaseChange{Device and  operation regime}}
\begin{figure*}[!t]
\centering
\includegraphics[width=0.94\linewidth]{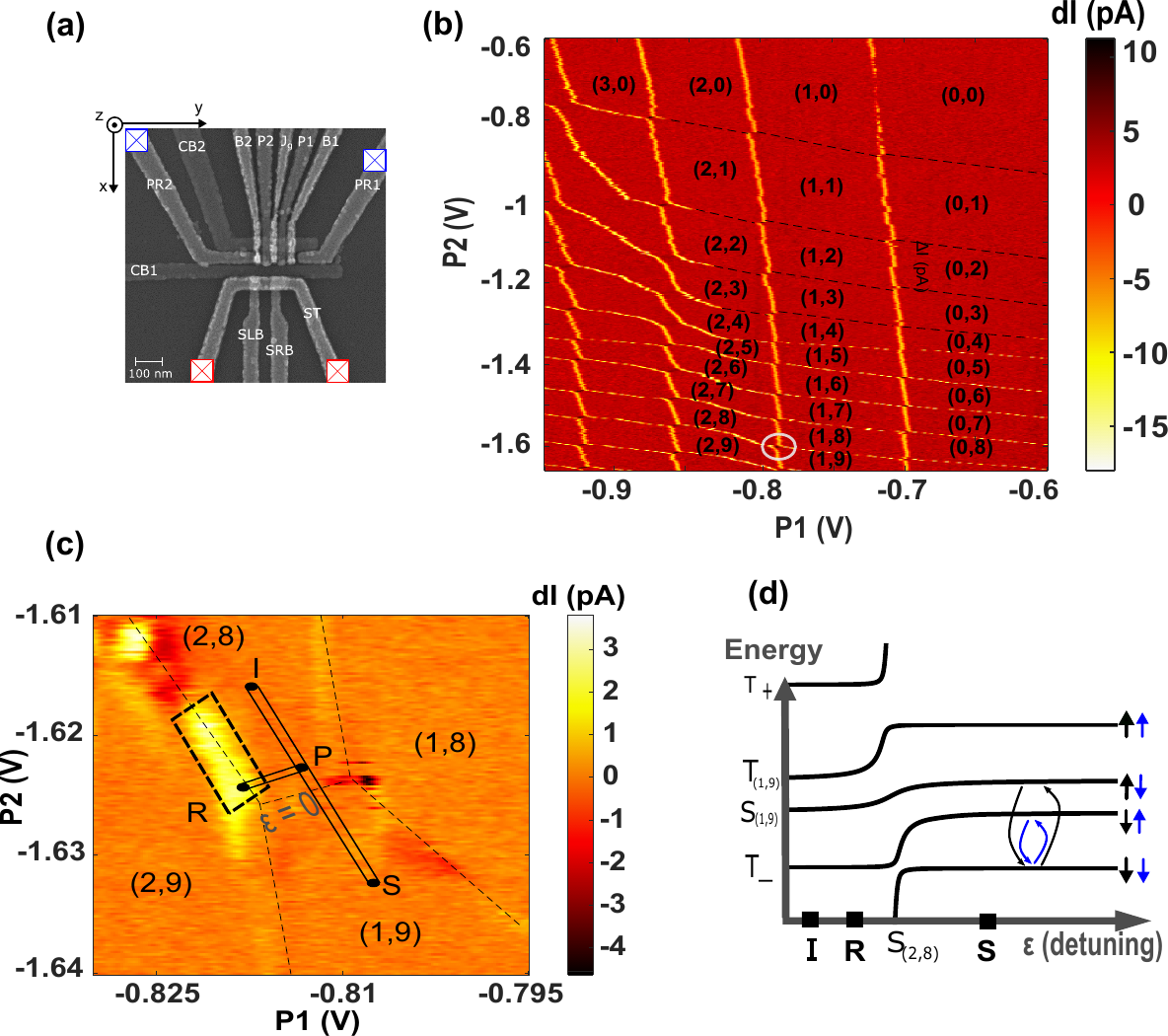}
\renewcommand{\figurename}{ FIG.}
\caption{Device operating point and latched readout. (a) SEM image of the device showing the layout of gates and the position of the quantum dot. Hole quantum dots are formed under gates P1 and P2. The difference in tunneling rates between the two dots (fast and slow) enables latched readout. (b) Charge stability diagram down to the last hole. The labels indicate the number of holes in the P1 and P2 quantum dots. The color axis represents the change in current through the sensor. The (2,8)$\leftrightarrow$(1,9) charge transition, marked with a white circle, corresponds to the operating regime. (c) (2,8)$\leftrightarrow$(1,9) charge transition regime. The pulse sequence starts with initialization (I) in (2,8) and then proceeds to S point which pushes the spin into a (1,1)-type configuration. For readout the pulse goes to PSB point (P) and then to R for latched readout. (d) Energy level diagram as a function of detuning $\epsilon$ illustrating the single spin states for the (2,8) $\rightarrow$(1,9) charge transition. Different colored arrow shows different possible single spin transitions.}
\label{figure_1_g-TMR_supp}
\end{figure*}

Fig.~\ref{figure_1_g-TMR_supp}(a) is the SEM image of the device. The device consists of single electron transistor (SET) defined by the top gate ST, and the barrier gates SLB and SRB. This SET is used for charge sensing. The region of the device defining hole quantum dots has been described in the main text.  
Fig.~\ref{figure_1_g-TMR_supp}(b) shows the charge stability diagram down to the last hole, measured using the charge sensor. The color scale corresponds to the change in the sensor current (dI) with respect to an applied square voltage pulse on the plunger gate P1~\cite{Elzerman_Charge_Sensing, Liles2024STQubit} with a repetition rate of $\SI{177}{\hertz}$. We monitor the charge sensor using lock-in detection referenced to the applied pulse and also maintain dynamic feedback on the SET. Using the charge sensor, we can determine the absolute occupation number of the hole quantum dots. We observe a loss of visibility for some transitions (black dashed lines) when the tunnel rate drops below the pulse cycling rate ($\SI{177}{\hertz}$), but we can infer all transitions from the inter-dot transitions. The region, (2,8)$\leftrightarrow$(1,9), marked with the white circle is used in the experiment. We use this region for our experiment as this is the first charge transition we encounter that shows Pauli Spin Blockade (PSB) and latched readout. For future work, we will also explore the other charge transitions and study the variability in the driving mechanism with a change in occupation number.
\begin{figure*}[!t]
\centering
\includegraphics[width=1\linewidth]{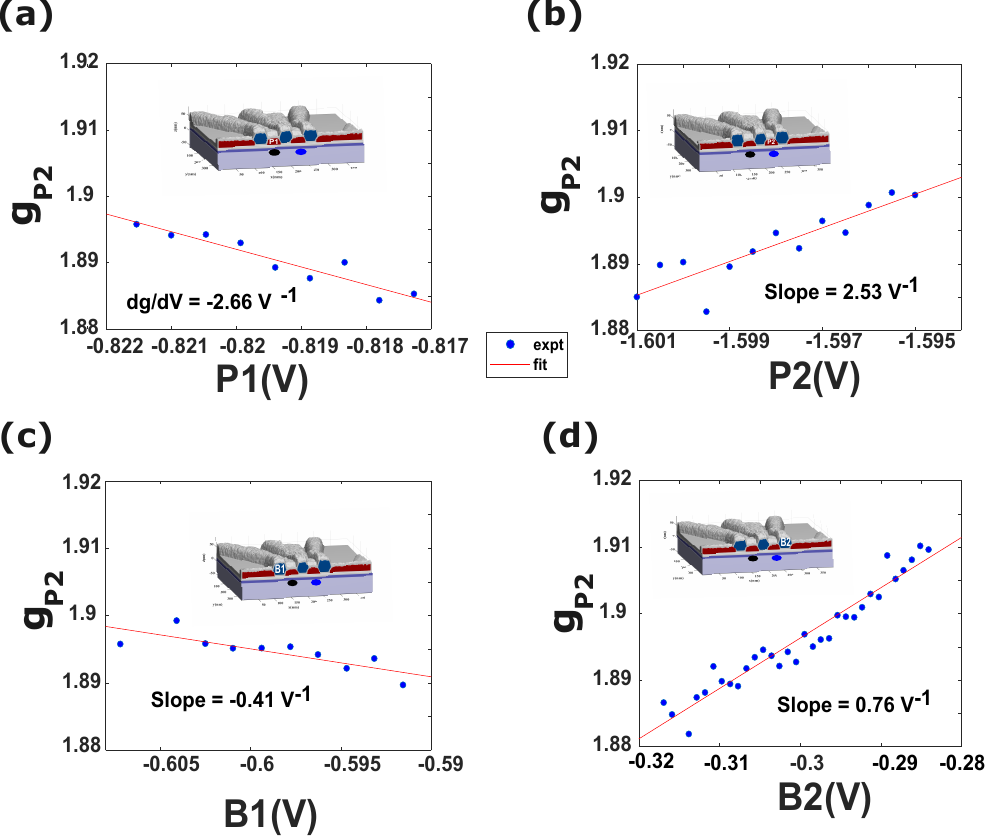}
\renewcommand{\figurename}{FIG.}
\caption{Gate dependence of $g$-factor. (a)-(b) $g$-factor is measured as a function of the plunger gates P1 and P2. The change in the $g$-factor is same for both the gates, indicating symmetrical control (given by slope).The inset is the device image with the relevant gate labeled. (c)-(d) Same experiment as (a)-(b) but with tunnel barrier gates B1 and B2. The $g$-factor change is twice as large for B2 compared to B1, indicating that the dot being driven is closer to B2.}
\label{figure_2_g-TMR_supp}
\end{figure*}
We perform all measurements in the (2,8)$\leftrightarrow$(1,9) charge transition regime which is equivalent to a (2,0)$\leftrightarrow$(1,1) spin system \cite{Liles2018}. Fig.~\ref{figure_1_g-TMR_supp}(c) is a zoom-in of (2,8)$\leftrightarrow$(1,9) with the pulse sequence (I-S-P-R) used for the latched readout. Black dashed lines mark the charge transition and inter-dot transition. The inter-dot transition line marks the zero de-tuning ($\epsilon$). The rectangular region indicated by the bold black dashed lines marks the latched region, which is used for performing spin to charge conversion. The color axis is the change in sensor current with respect to the applied pulse shown in the figure and a reference pulse (S-I-P-R).

The pulse starts from position I, which is used for initialization in the S(2,8) state. This is achieved by dwelling at I for $\SI{500}{\nano\second}$. Moving along the detuning axis ($\epsilon$), the pulse goes to point S in (1,9) regime, which pushes one of the spins from the left dot to the right dot, forming a (1,1)-like system. At point S, the spins are allowed to evolve. Depending on the experiment, the dwell time at point S is varied. Point P is a transitory point with a short dwelling time ($\SI{1}{\nano\second}$) to avoid any unwanted charge transition occurring. From point P we go to the readout point R performing latched PSB readout. The dwelling time at R was $\SI{2}{\micro\second}$.   

Fig.~\ref{figure_1_g-TMR_supp}(d) shows the energy spectrum of a double quantum dot in a finite magnetic field as a function of $\epsilon$, which corresponds to the (2,0)$\leftrightarrow$(1,1) spin transitions. At large $\epsilon$, the quantum dots are effectively decoupled due to the reduced wavefunction overlap, resulting in a negligible exchange interaction ($J \approx 0$). In this regime, the holes are well localized in their respective dots, and the system can be treated as two individually controllable spins.
\begin{figure*}
\centering
\includegraphics[width=0.8\linewidth]{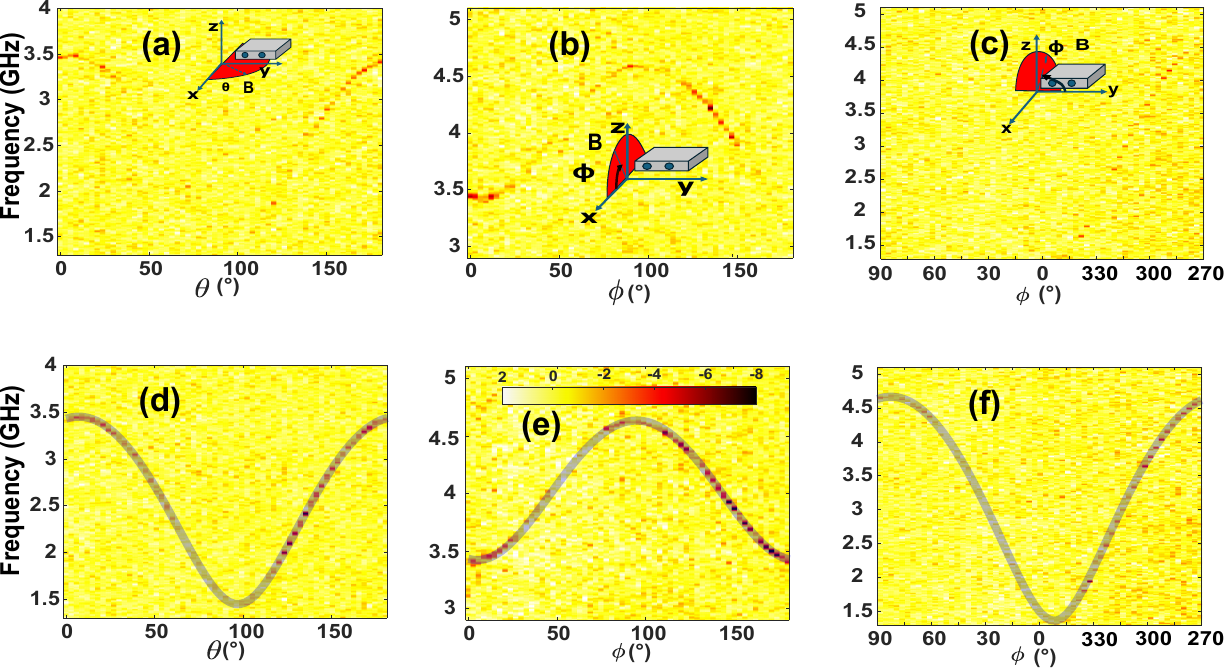}
\renewcommand{\figurename}{FIG.}
\caption{EDSR rotation. (a)-(c) MW frequency with change in orientation of magnetic field at MW power of $-13$dBm. The inset gives the plane of rotation. (d)-(f) Same experiment as (a)-(c) with MW power set at $-10$ dBm. The continuous line is a fit to the equation (1) given in the main text. The magnetic field was set constant magnitude of 0.13 T in all of the experiments. The voltage pulse at the plunger gate P1 is $0.02$ V }
\label{figure_3_g-TMR_supp}
\end{figure*}

\section{\NoCaseChange{Dot location via \textit{g}-factor response}}
\begin{figure*}[!t]
    \centering
    \includegraphics[width=1\linewidth]{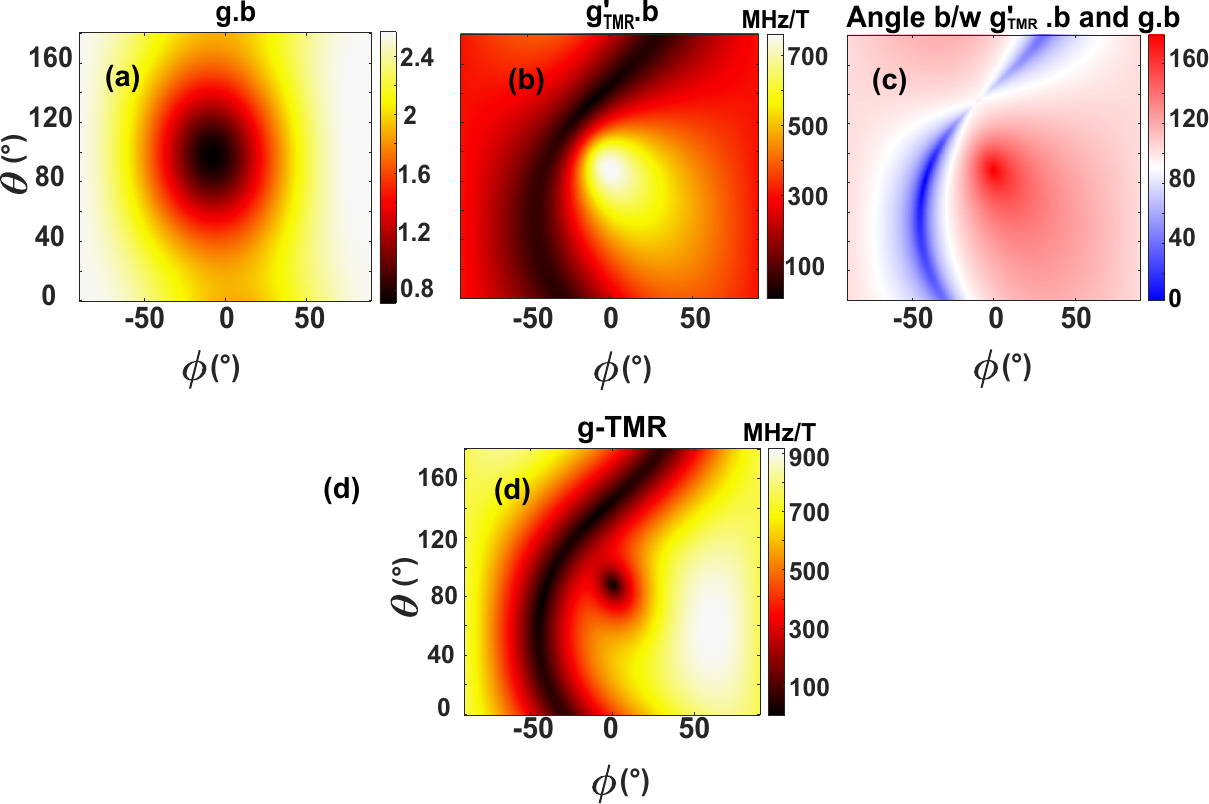}
    \renewcommand{\figurename}{FIG.}
    \caption{$g$-TMR. (a) $\hat{g}\cdot\hat{b}$ computed from Rabi equation for different magnetic field orientations. The color scale is g-factor (b) $\hat{g}'_\mathrm{TMR}\cdot\hat{b}$ computed for different magnetic field orientations. The color scale is Rabi frequency in MHz/T. (c) Angle between the $\hat{g}'_\mathrm{TMR}\cdot\hat{b}$ and $\hat{g}\cdot\hat{b}$. The color scale is degree. (d) $g$-TMR contribution to the spin driving for different magnetic field orientations.}
    \label{figure_4_g-TMR_supp}
 \end{figure*}

We next perform an investigation to determine which quantum dot hosts the qubit being addressed. We measured the $g$-factor as a function of the voltages applied to the plunger and barrier gates (device figure given in the inset of Fig.~\ref{figure_2_g-TMR_supp}). 
Figs.~\ref{figure_2_g-TMR_supp}(a)-(b) show the $g$-factor measured for different voltages of the plunger gates P1 and P2, respectively. The slope of each line ($\frac{\partial g}{\partial V_\mathrm{P1}}$ and $\frac{\partial g}{\partial V_\mathrm{P2}}$) is similar, which suggests that both plunger gates influence the dot equally. However, despite the similar plunger gate influence, only a single EDSR line is observed. Next, we investigate the response of the tunnel barriers as shown in Fig.~\ref{figure_2_g-TMR_supp}(c)-(d). The slope of B2 is twice that of B1, indicating a stronger influence of gate B2 on the $g$-factor. This likely reflects the proximity of the dot to B2. We attribute the presence of a single visible line to spin rotations occurring in the dot closer to P2, which hosts more holes than the dot under gate P1. To accommodate more holes, the dot under P2 is larger, increasing the spatial extent of the wavefunction and thereby enhancing the gate lever arm, which strengthens the microwave coupling. In the $g$-matrix modulation analysis considered here, the extracted parameters do not depend on whether dot under P1 or P2 is driven. However, for future experiments designed to selectively probe different spin transitions, it may be necessary to choose which dot occupancy to modulate.
\section{\NoCaseChange{\textit{g}-TMR and IZ from \textit{g}-matrix formalism}}
Rabi frequency is given by the expression \cite{Crippa2018,Ares2013} 

\begin{equation}
f_\mathrm{Rabi}=\frac{\mu_\mathrm{B} B V_\mathrm{ac}}{2h|g^*|}[\hat{g}(V_0) \cdot \hat{b}] \times [\hat{g'}(V_0) \cdot \hat{b}],
\label{equation_S1}
\end{equation}
where $\hat{g}'$ is the $g$-matrix derivative with respect to the gate voltage, $\textbf{b}$ is the unit vector along the direction of the magnetic field, and $V_\mathrm{ac}$ is the amplitude of the driving signal.
To disentangle the individual contributions of spin driving, we express the derivative of the $g$-matrix, $\hat g'$ as the sum 
\begin{equation}
    \hat{g}'=\hat{g}'_\mathrm{TMR} + \hat{g}'_\mathrm{IZ},
\label{equation_S2}
\end{equation}
where $\hat{g}'_\mathrm{TMR}$ and $\hat{g}'_\mathrm{IZ}$ can be calculated by decomposing $\hat{g}^T\cdot\hat{g}'$ into a symmetric ($\hat{S}$) and an antisymmetric matrix ($\hat{A}$):
\begin{equation}
    \hat{g}^T\cdot\hat{g}' = \hat{S} + \hat{A}.
\label{equation_S3}
\end{equation}
$\hat{g}'_\mathrm{TMR}$ can be determined by the expression\cite{Crippa2018}:
\begin{equation}
    \hat{g}'_\mathrm{TMR} = (\hat{g}^{-1})^T\cdot\frac{\hat{G}'}{2}.
\label{equation_S4}
\end{equation}

Using $\hat{G'}$ given in the main text, we can determine the contribution of $\hat{g}'_\mathrm{TMR}$. $\hat{g}'_\mathrm{IZ}$ is independent of $\hat{G'}$\cite{Crippa2018}, but it is related to $\hat{g}$:
\begin{equation}
    \hat{g}'_\mathrm{IZR} = (\hat{g}^{-1})^T\cdot{\hat{A}}.
\label{equation_S5}
\end{equation}
Since our measurements are performed in the lab frame with non-diagonal $\hat{g}$, it is instructive to define $\hat{g}_\mathrm{IZR}$ in a basis in which $\hat{g}$ is diagonal and rotate $\hat{g}'_\mathrm{IZR}$ back to the lab frame. The rotation matrix from the lab frame to the frame in which $\hat{g}$ is diagonal is the matrix obtained from the eigenvectors of $\hat{G}$.

The three independent elements of $\hat{A}$ and $V_\mathrm{ac}$ are treated as free parameters and fitted to the experimentally determined Rabi frequencies using Eq.~\ref{equation_S1}. The fit estimates $V_\mathrm{ac} = \SI{0.002}{\volt}$. The $\hat{g}'_\mathrm{IZR}$ matrix is given in Eq.~6 in the main text.

\section{EDSR}
Fig.~\ref{figure_3_g-TMR_supp} shows the sensor current response measured as a function of microwave frequency and magnetic field orientation, with microwave power set to $\SI{-13}{dBm}$ (Figs.~\ref{figure_3_g-TMR_supp}(a)-(c)) and $\SI{-10}{dBm}$ (Figs.~\ref{figure_3_g-TMR_supp}(d)-(f)). These values are the source power levels before attenuation. In Fig.~\ref{figure_3_g-TMR_supp}(a), we observe the disappearance of the EDSR signal as we move away from the x-direction. This disappearance of EDSR explains the absence of Rabi oscillations beyond $\theta = \SI{30}{\degree}$ in Fig.~\ref{figure_3_g-TMR_supp}(b) of the main text. In the angular range between $\theta = \SI{30}{\degree}$ and $\SI{147}{\degree}$, no EDSR signal is visible. With an increase in microwave power (Fig.~\ref{figure_3_g-TMR_supp}(d)), we see the appearance of EDSR signal for some of the magnetic field orientations, but there are still orientations that show no EDSR signal. The same behavior is found comparing Fig.~\ref{figure_3_g-TMR_supp}(e) to Fig.~\ref{figure_3_g-TMR_supp}(b) and Fig.~\ref{figure_3_g-TMR_supp}(f) to Fig.~\ref{figure_3_g-TMR_supp}(c). This set of figures emphasize that the driving efficiency depends on the driving power, but it is not the only factor that affects driving. In the main text (Fig.~2(d)), we show that the orientation where the electrical noise is dominant corresponds to the orientation where the EDSR signal is not visible. The transparent line in Figs.~\ref{figure_3_g-TMR_supp}(d)-(f) is fitted to Eq.~1 in the main text for extracting $\hat{G}$.
 
\section{\NoCaseChange{\textit{g}-TMR}}
Supplementary Fig.~\ref{figure_4_g-TMR_supp} is the extended dataset for Fig.~\ref{figure_2_g-TMR_supp}(a) in the main text. In these figures, we plot the individual components of the Rabi equation, $\hat{g}\cdot\hat{b}$ and $\hat{g}'_\mathrm{TMR}\cdot\hat{b}$, and the angle between them to know which term in the Rabi equation influences the $g$-TMR contribution. Supplementary Fig.~\ref{figure_4_g-TMR_supp}(a) is $g\cdot b$ computed from the Rabi equation. The color scale is the magnitude of the $g\cdot b$, which is minimized in the y-direction and maximized as we move out-of-plane ($\phi$). 

Fig.~\ref{figure_4_g-TMR_supp}(b) shows the variation of term $\hat{g}'_\mathrm{TMR}\cdot\hat{b}$ for different magnetic field orientations. Since the Rabi frequency is also influenced by the angle between the $\hat{g}'\cdot\hat{b}$ and $\hat{g}\cdot\hat{b}$ in Fig.~\ref{figure_4_g-TMR_supp}(c) we compute the angle between the $\hat{g}'_\mathrm{TMR}\cdot\hat{b}$ and $\hat{g}\cdot\hat{b}$. Fig.~\ref{figure_4_g-TMR_supp}(d) gives the $g$-TMR contribution to the spin driving for different magnetic field orientations. Using Figs.~\ref{figure_4_g-TMR_supp}(a)-(c) we conclude that the shape of the $g$-TMR contribution is primarily influenced by the $\hat{g}'_\mathrm{TMR}\cdot\hat{b}$ term. The minima in Fig.~\ref{figure_4_g-TMR_supp}(d) follow the same shape as those in Fig.~\ref{figure_4_g-TMR_supp}(b). We can also explain the small island of minima in $g$-TMR seen in Fig.~\ref{figure_4_g-TMR_supp}(d), which is a consequence of the angle being $\SI{180}{\degree}$ between the $\hat{g}'_\mathrm{TMR}\cdot\hat{b}$ and $\hat{g}\cdot\hat{b}$, even though the $\hat{g}'_\mathrm{TMR}\cdot\hat{b}$ term is maximal.

\bibliographystyle{apsrev4-2}
\bibliography{sample}
\end{document}